\documentstyle{article}
\begin{document}
\renewcommand{\theequation}{\thesection.\arabic{equation}}
\parskip5pt
\title
{NONLINEAR INTEGRABLE SYSTEMS
RELATED TO
ARBITRARY SPACE-TIME DEPENDENCE
OF THE SPECTRAL TRANSFORM}
\author{J\'er\^ome LEON\\
Physique Math\'ematique et Th\'eorique, CNRS-URA 768\\
Universit\'e Montpellier II, 34095 MONTPELLIER FRANCE}
\date{ }
\maketitle

\begin{abstract}
We propose a general {\em algebraic analytic} scheme for the spectral transform
of
solutions of nonlinear evolution equations. This allows us to give the general
integrable evolution corresponding to an arbitrary time and space dependence of
the spectral transform (in general nonlinear and with non-analytic dispersion
relations). The main theorem is that the compatiblity conditions gives always a
true nonlinear evolution because it can always be written as an identity
between
polynomials in the spectral variable $k$. This general result is then used to
obtain first a method to generate a new class of solutions to the nonlinear
Schr\"odinger equation, and second to construct the spectral transform theory
for
solving initial-boundary value problems for resonant wave-coupling processes
(like self-induced transparency in two-level media, or stimulated Brillouin
scattering of plasma waves or else
stimulated Raman scattering in nonlinear optics etc...).
\end{abstract}
\vskip1in
Preprint PM94/01 \hskip2in PACS \# 02.30 Jr, 03.40 Kf
\vfill\eject
\section{Introduction}

The spectral transform method to build and solve classes of nonlinear
evolution equations (for a field $Q(x,t))$ works as a nonlinear
extension
of the Fourier transform \cite{akns} : it associates to the field
$Q(x,t)$ its spectral transform $R(k,x,t)$ which is a distribution in
the spectral complex variable k and a function of the real space and
time variables $x$ and $t$. The domain of definition for $k$ and the
$(x,t)$-dependence of $R$ uniquely determine both the related nonlinear
evolution for $Q$ and the class of functions to which it belongs.

For instance when $k$ varies on ${\bf C}$ and when $R$ is a $2 \times 2$
off-diagonal matrix depending on $x$ and $t$ through
\begin{equation}R_x = [R,ik \sigma_3],\hskip10pt R_t =[R, ik^2
\sigma_3], \label{evonls}\end{equation}
then $Q$ is a $2\times 2$ off-diagonal matrix obeying the nonlinear
Schr\"odinger equation
\begin{equation} i \sigma_3 Q_t = -{1\over 2} Q_{xx} + Q^3.
\label{matnls}\end{equation}
This is the very reason why the spectral transform allows us to solve
the nonlinear evolution (\ref{matnls}) : the corresponding evolution
in the spectral space is linear. It is then enough to have a bijection
between $Q$ and $R$, which is the set of direct and inverse
spectral problems.

The main purpose of this work is to answer the following simple
question:
can we still build and solve nonlinear evolutions when the
$(x,t)$-dependence of the spectral transform is {\em arbitrary}. We
shall
prove in particular that the nonlinear evolutions associated with a
local $2\times 2$ matrix $\bar{\partial}$ problem on the Riemann sphere
can always be written in closed form whatever be the space and time
dependences of $R$ in a very general class of equations (not necessarily
linear).

Here the $\bar{\partial}$ problem will be understood as being the main
tool which links $Q(x,t)$ to $R(k,x,t)$ by means of a matrix valued
function $\mu (k,x,t)$ solution of
\begin{equation}\mu(k) = {\bf 1} + {1\over 2\pi} \int\!\!\!\int
{d\lambda\wedge d\bar{\lambda}\over \lambda - k} \mu (\lambda)
R(\lambda)\label{cauchy}\end{equation}
(being parametrically induced by
$R(k,x,t)$, the $(x,t)$-dependences can be omitted everywhere in $\mu$
and $R$).
If we operate with the $\bar\partial$-operator (see next section) on
both side of
the above equation we obtain the so-called $\bar{\partial}$ problem for
$\mu$ :
\begin{equation} {\partial\over\partial\bar k} \mu (k) = \mu(k) R(k) ,
\hskip10pt\mu(k)\displaystyle{\mathop{\to}_{k\to\infty}} {\bf 1}
\label{dbar}\end{equation}
It is clear that any dependence of $R(k)$ on external
{\em independent} (real) parameters
($x$ and $t$) induces a dependence of $\mu(k)$ generally through an
{\em overdetermined system} of differential equations : the so-called
Lax pair. Then this overdetermination implies some constraints on the
coefficients of the Laurent series for $\mu(k)$ (see (\ref{serie})
next section).
When they form a closed system independent of the variable
$k$, these constraints are nothing but the nonlinear evolution equation
related to the given space and time dependence of $R(k)$.

The result presented here will undercover some interesting properties of
overdetermined systems (for $R(k)$ and for $\mu(k)$) when they are
related through a $\bar{\partial}$ problem like (\ref{dbar}). First
we
shall discover that the differential system for $R$ (data of $R_t$ and
$R_x$) although very general, has to be written in a special form in
order to obtain a non-trivial integrable related evolution. This remark
will actually bring into light the necessity of a {\em spectral
parameter} in the Lax pair. Second, our result will be shown to generate
new classes of integrable systems with in particular some very
interesting applications to problems of interaction of radiation with
matter (for instance stimulated Brillouin and Raman scattering), which
are always associated with a mixed initial and boundary value problem.

Although some subcases and applications of the material presented here have
been
reported elsewhere in \cite{jla}...\cite{jle}, the results presented here are
mainly new and we shall avoid
using directly any previous result. In that purpose, the next section is
a brief recall of the basic mathematical tools that we have chosen to
present in their {\em light} version, reporting to \cite{MUSKH} for
mathematical details.

The section 3 is the heart of this paper : the main theorem is
established and we obtain the general {\em algebraic-analytic} scheme of the
spectral transform method.
Some important consequences of general interest are then
discussed.

In section 4 we present an application of the general theory to the nonlinear
Schr\"odinger equation (NLS) and exhibit in particular a new method to
generate a more general class of solutions.

The section 5 is devoted to the main interesting application of the general
result:
the theory is used to build integrable coupled systems
for which some arbitrary boundary values are prescribed. As a consequence the
time evolution of the spectral transform is in general nonlinear with some
very new properties of the solution in physically interesting cases.

\vfill\eject

\setcounter{equation}{0}
\section{Notations, definitions and basic tools}

The derivative with respect to $\bar{k}$,($k = k_R +ik_I$) formally
defined as
\begin{equation}{\partial\over \partial\bar{k}} = {1\over 2} \left (
{\partial\over \partial k_R} + i{\partial\over \partial k_I}\right )
\end{equation}
vanishes on the space of analytic functions of $k$ (it is sometimes
called the measure of non-analyticity). It can be conveniently defined
by the Cauchy-Green representation of a function in $\bf C$:
\begin{eqnarray} f(k)& = & f_0(k) + f_1(k)\nonumber\\
f_1(k)& = & {1\over 2i\pi} \int\!\!\!\int {d\lambda\wedge
d\bar{\lambda}\over
\lambda - k} {\partial f(\lambda)\over \partial\bar{\lambda}}
\label{green}\end{eqnarray}
where $f_0(k)$ is the asymptotic behavior (analytic in $k$) of $f(k)$
when $k\rightarrow \infty$.\par
Here we shall work more particularily in the space of function $f(k)$
whose behavior
$f_0(k)$ is a polynomial (in $k$) which we note ${\cal P}_k$, that is
\begin{equation}f_0(k) = {\cal P}_k\{f\} \label{defpol}
\end{equation}
We call ${\cal H}_p$ the space of functions which have the
representation (\ref{green}) with the constraint (\ref{defpol}).

A consequence of (\ref{green}) is that we can formally write
\begin{equation} {\partial\over \partial\bar{k}} {1\over \lambda - k} =
2i\pi \delta(\lambda-k) \label{dirac} \end{equation}
where $\delta$ is the Dirac distribution in $\bf C$, namely
$$ \int\!\!\!\int d\lambda\wedge d\bar{\lambda} f(\lambda)
\delta (\lambda - k) = f(k)$$
\begin{equation}\equiv -2i\int^{+\infty}_{-\infty} d\lambda_R
\int^{+\infty}_{-\infty}
d\lambda_I f(\lambda_R + i\lambda_I) {i\over 2} \delta (\lambda_R -k_R)
\delta (\lambda_I - k_I). \label{distrib}\end{equation}
We shall also use the distributions $\delta^{\pm}$ defined by
\begin{equation} \int\!\!\!\int d\lambda \wedge d\bar{\lambda} f(\lambda)
\delta^{\pm} (\lambda_I) = - 2i \int^{+\infty}_{-\infty} d\lambda_R
f(\lambda_R \pm i0), \label{deltapm} \end{equation}
and make use of the notation
\begin{equation}f^\pm(k_R)=f(k_R \pm i0).\label{notepm} \end{equation}
{}From (\ref{dirac}) we can deduce
\begin{equation} {\partial\over \partial\bar{k}}
\int^{+\infty}_{-\infty} {d\lambda_R\over \lambda_R - k \pm i0}
f(\lambda_R) = - \pi f(k) \delta^{\pm} (k_I). \label{distripm}
\end{equation}
Let us recall also the Sokhotski-Plemelj formula
\begin{equation}
\int^{+\infty}_{-\infty} {d\lambda\over \lambda - k \pm i0} f(\lambda) =
\pm i\pi f(k) + P\!\!\!\int^{+\infty}_{-\infty} {d\lambda\over
\lambda-k} f(\lambda) , \label{ptem} \end{equation}
where$P\!\!\!\int$ stands for the Cauchy principal value integral.
The two above formulae allow us to obtain by direct calculation
\begin{equation} {i\over 2} \left [ f(k) \delta^+(k_I) - f(k)
\delta^-(k_I)\right ] = {\partial\over \partial\bar{k}} f(k),
\label{discout}\end{equation}
which provides the $\bar{\partial}$ derivative of a function
discontinuous on the real axis but analytic elsewhere.

We note that the Cauchy-Green integral equation (\ref{green})
can also be used to define the {\em inverse} of the $\bar{\partial}$
operator in ${\cal H}_p$ as
\begin{equation} \bar{\partial}^{-1} \left ( {\partial f(k)\over
\partial \bar{k}} \right ) = f(k) - {\cal P}_k\{f\},\label{invdbar}
\end{equation}
which is nothing but $f_1(k)$. Hence the k-polynomial part of $f(k)$
appears as the constant of integration in ${\cal H}_p$ of the
$\bar{\partial}$ operator.
Whenever possible we use the short-hand notation
\begin{equation} \bar{\partial}f = {\partial\over \partial\bar{k}} f(k),
\hskip10pt
\bar{\partial}^{-1}f={1\over 2i\pi} \int\!\!\!\int {d\lambda\wedge
d\bar{\lambda}\over \lambda - k} f(\lambda) \label{notdbar}
\end{equation}
and we have the very important property
\begin{equation} \bar{\partial}^{-1} f = O({1\over k})\ \mbox{as}\ \
k\rightarrow \infty \label{behav} \end{equation}

Next we adopt for the Laurent series of $f(k)$ the following notation
\begin{equation} f(k) = f_0(k) + \sum^{\infty}_{n=1} k^{-n} f^{(n)} ,
\label{laurent} \end{equation}
where from (\ref{green})
\begin{equation} f^{(n)} = {-1\over 2i\pi} \int\!\!\!\int d\lambda\wedge
d\bar{\lambda}\;\; {\partial f(\lambda)\over \partial \bar{\lambda}}
\;\;\lambda^{n-1} \label{serie} \end{equation}

 For a complex valued
function $f(k)=a(k)+ib(k)$ of the complex variable $k=k_R+ik_I$, we
note
\begin{equation}
\bar f(k)=a(k_R+ik_I)-ib(k_R+ik_I),\label{conjug}\end{equation}
\begin{equation}
f^*(k)=a(k_R-ik_I)-ib(k_R-ik_I)=\bar f(\bar k).\label{star}
\end{equation}
Then, from the definition  (\ref{deltapm}), since $\delta^\pm$ are real valued
distributions obeying $\delta^+(k_I)=\delta^-(-k_I)$, we have
\begin{equation}
\delta^+(k_I)=(\delta^-(k_I))^*.\label{deltaconj}\end{equation}

It is also usefull to remember that, when $\lambda$ depends on a
real parameter, say $t$, we may write
\begin{equation}{\partial\over \partial t} \delta (k-\lambda) =
{\delta (k-\lambda)\over k-\lambda}
{\partial \lambda\over \partial t} ,
\label{derivdelta}\end{equation}
which is meaningfull on the space of functions of $k$ that vanish in
$k=\lambda$.

Finally, we shall work throughout the paper in the space of complex-valued
$2\times2$ matrices, say $M$, which elements $m_{ij}$ are differentiable
functions
of two real variables ($x$ and $t$) and distributions in a complex variable
($k$).
It is convenient to split the matrices in their diagonal and anti-diagonal
parts,
namely
\begin{equation}M=M^D+M^A,\hskip10pt
M^D=\left(\matrix{m_{11} & 0\cr0 & m_{22}}\right),\hskip10pt
M^A=\left(\matrix{0 & m_{12} \cr m_{21}& 0}\right).
\label{offdiag}\end{equation}
Using the standard definition of the Pauli matrices
\begin{equation}{\bf 1} = \left (\matrix {1&0\cr 0&1}\right ), \sigma_1
= \left (\matrix {0&1 \cr 1&0}\right ), \sigma_2 = \left ( \matrix
{0&-i\cr i&0}\right ), \sigma_3 = \left (\matrix {1&0\cr 0 & -1}\right )
\label{pauli}\end{equation}
we have for instance
\begin{equation}[\sigma_3, M]=2\sigma_3 M^A.\label{commutoff}\end{equation}

\vfill\eject

\setcounter{equation}{0}
\section{General theorem}

The starting point is the Cauchy-Green integral equation
(\ref{cauchy})
or equivalently the $\bar{\partial}$-problem (\ref{dbar}) for the
matrix valued complex function $\mu(k)$ :
\begin{equation} {\partial\over \partial\bar{k}} \mu(k) = \mu(k) R(k) ,
\hskip10pt
\mu(k) = {\bf 1} + O({1\over k}), \label{start} \end{equation}
where the given datum $R(k)$ is a distribution in the set of $2\times 2$
matrices of vanishing trace:
\begin{equation} tr\{R(k)\}=0.\label{roff}
\end{equation}

{\bf Theorem 1:}
\begin{center}
\fbox%
{%
\begin{minipage}{11.5cm}
When $R(k)$ depends on two real parameters $x$ and $t$ according to
\begin{equation} R_x = [R,\Lambda] + N , \label{rx}\end{equation}
\begin{equation} R_t = [R, \Omega] + M, \label{rt} \end{equation}
the quantities
\begin{equation} U_0 = - {\cal P}_k\{\mu\Lambda\mu^{-1}\},\hskip10pt
U_1 = \bar{\partial}^{-1} \{\mu (N-
\bar{\partial}\Lambda) \mu^{-1}\} \label{ulax} \end{equation}
\begin{equation} V_0=-{\cal P}_k\{\mu \Omega \mu^{-1}\},\hskip10pt
 V_1 = \bar{\partial}^{-1} \{\mu
(M-\bar{\partial}\Omega) \mu^{-1} \} \label{vlax} \end{equation}
obey the following equation
\begin{equation} U_{0,t}-V_{0,x}+[U_0,V_0]+{\cal P}_k\{\mu A\mu^{-1}\}=
{\cal P}_k \{[V_0,U_1] - [U_0, V_1]\},\label{main}\end{equation}
where the matrix $A$ stands for
\begin{equation} A = \Lambda_t - \Omega_x + [\Omega, \Lambda],
\label{adisp} \end{equation}
and where the entries $\{\Lambda,N,\Omega,M\}$ obey the compatibility
constraint
\begin{equation}N_t -[N,\Omega] - M_x + [M,\Lambda] = [A,R].
\label{const} \end{equation}
\end{minipage}}\end{center}

Equation (\ref{main}) is a {\em nonlinear evolution equation}
because it is an identity between
{\em polynomials} in $k$ and then it has to be read as a set of local
$k$-independent differential equations (each coefficient of a given
power of $k$ vanishes).

Here above the $2\times 2$ matrices $\Lambda$ and $\Omega$ are
functions
in ${\cal H}_p$ (see sec. 2) and $M$ and $N$ are in general
distributions. All
these 4 given matrices are differentiable fuctions of the real variables
$x$ and $t$.
Up to now the $(x,t)$-dependence (\ref{rx})(\ref{rt}) of $R$ is
quite arbitrary but we will
understand later why this specific structure is important (rather than
just letting $\Lambda = \Omega = 0$). Note also that
nothing prevents these equations to be {\em nonlinear} in $R$.

The problem is to find the $(x,t)$-dependence of the solution $\mu(k)$
of (\ref{start}) when $R(k)$ is required to solve (\ref{rx}) and
(\ref{rt}). The method proposed in \cite{jlc} consists in using first
the commutativity of the differential operators.
\begin{equation} \left [ {\partial\over \partial t},
{\partial\over \partial\bar{k}} \right ] = 0 ,\hskip10pt
\left [ {\partial\over \partial x} , {\partial\over
\partial\bar{k}} \right ] = 0 ,  \label{comm} \end{equation}
and second the property that $\mu(k)$ is a regular matrix.
Indeed from (\ref{roff}) we easily obtain
\begin{equation} {\partial\over \partial\bar{k}} det \mu(k) = 0
\end{equation}
and hence, since $\mu(k) = {\bf 1} + 0 ({1\over k})$ for large $k$, the
Liouville theorem implies
\begin{equation} det \mu(k) = 1 \label{det}\end{equation}

These results allow us to obtain by direct computation the following
relations
\begin{equation} \bar{\partial} \{(\mu_x - \mu\Lambda) \mu^{-1} \} = \mu
(N - \bar{\partial}\Lambda) \mu^{-1} \end{equation}
\begin{equation} \bar{\partial} \{(\mu_t - \mu\Omega) \mu^{-1} \} = \mu
(M - \bar{\partial}\Omega) \mu^{-1} \end{equation}
They can be integrated by using (\ref{invdbar}), the result can be
written as the following overdetermined system
\begin{equation} \mu_x =\mu\Lambda + U\mu\ ,\hskip10pt \mu_t = \mu\Omega
+V\mu, \label{lax}\end{equation}
in which $U$ and $V$ are in ${\cal H}_p$, that is
\begin{equation} U = U_0 + U_1\ ,\ V = V_0 + V_1,
\label{uv}\end{equation}
where $U_0, U_1, V_0, V_1$ are given in (\ref{ulax}) and (\ref{vlax}).

Consequently the $\bar{\partial}$ problem (\ref{start}) serves as the
link between the two sets of overdetermined differential systems
(\ref{rx}) (\ref{rt}) and (\ref{lax}). Computing $R_{xt}$ in two ways
leads to the constraint (\ref{const}) and doing the same for
$\mu_{xt}$  leads to the compatibility condition:
\begin{equation} U_t - V_t + [U,V] = - \mu A \mu^{-1}. \label{curva}
\end{equation}

We come now to the demonstration of (\ref{main}). This is done first
by
computing the quantity $U_t - V_x$ appearing in (\ref{curva}) with
the definitions (\ref{ulax}) to (\ref{vlax}). To arrange
conveniently
this quantity we have to use the inverse of (\ref{ulax}) and
(\ref{vlax}) that is
\begin{equation} \bar{\partial} U = \mu (N-\bar{\partial} \Lambda)
\mu^{-1}\ \ ,\ \ \bar{\partial} V = \mu (M-\bar{\partial} \Omega)
\mu^{-1} \end{equation}
with this help one can write
\begin{eqnarray} U_{1,t} - V_{1,x} & = & \bar{\partial}^{-1}\{
\mu(N_t-M_x -[N,\Omega] + [M,\Lambda]) \mu^{-1}-\nonumber\\
& - & \mu\bar{\partial} (\Lambda_t-\Omega_x + [\Omega,\Lambda])
\mu^{-1}- [U,\bar{\partial}V] - [\bar{\partial}U,V] \}.
\end{eqnarray}

At this stage enters the constraint (\ref{const}) which guarantees
the compatibility of the choices for $R_x$ and $R_t$. Then the above
equation can be written
\begin{equation} U_{1,t} - V_{1,x} = -\bar{\partial}^{-1} \{\mu
(\bar{\partial} A + [R,A]) \mu^{-1} + \bar{\partial} [U,V]\}
\end{equation}
or else
\begin{equation} U_{1,t} - V_{1,x} = - \bar{\partial}^{-1}
\bar{\partial}  \{ \mu A\mu^{-1} + [U,V]\} \end{equation}

The inversion formula (\ref{invdbar}) then gives
\begin{equation} U_{1,t} - V_{1,x} = - \mu A\mu^{-1}-[U,V]+{\cal P}_k
\{\mu A \mu^{-1} + [U,V]\}\end{equation}
which then, inserted in (\ref{curva}), easily leads to
(\ref{main}) fo we have obviously
\begin{equation}
{\cal P}_k\{ [U,V]\}=[U_0,V_0]+
{\cal P}_k \{ [U_0, V_1]+[U_1,V_0] \}.\end{equation}

First we note that this result is far from being trivial. Indeed, looking
at the structure of $U$ and $V$ given in (\ref{ulax}) and (\ref{vlax}),
we realize that the equation (\ref{curva}) involves quite
complicated functions of $k$ through many $\bar{\partial}^{-1}$ operators (see
(\ref{notdbar})). The relative simplicity of the result (\ref{main})
originates from the fact that all terms in the integral part of (\ref{curva})
can actually be written as an exact $\bar\partial$-derivative.

Second it is the very reason why the spectral transform works also in
the case of {\em singular despersion relation}. In previous works (f.i.
in \cite{jla} to \cite{jle}) the $k$-dependence in (\ref{curva}) was
seen to cancel out as a result of the {\em particular} choice of $\Lambda$ but
in no way as a very {\em general} property.

Third the result (\ref{main}) justifies the choice of the structure
in (\ref{rx}) (\ref{rt}). Indeed, avoiding any structure in
(\ref{rx}) and (\ref{rt}) by just letting $\Omega = \Lambda =0$,
makes the equation (\ref{main}) {\em trivially solved}: we
would have from (\ref{ulax}) and (\ref{vlax}) $U_0=0, V_0=0$ and hence
(\ref{curva}) would not lead to any nonlinear evolution equation.

Note also that another necessary condition that (\ref{curva}) be a
nonlinear evolution equation is that $U_0$ be nonzero, that is from
(\ref{ulax}) and (\ref{start}) that ${\cal P}_k \{\Lambda\}$
be non zero.
This property can be understood as the necessity of a {\em spectral
parameter} in the spectral problem (\ref{lax}) (here ${\cal P}_k \{\Lambda\}$
would play the role of the spectral parameter).

We have finally built a set of relations between partial differential equations
(obtained by means of the $\bar\partial$-problem) which has to be
understood as follows: if the set $\{R,\Omega,\Lambda,M,N\}$
obeys the system (\ref{rx})(\ref{rt}) and the compatibility
constraint (\ref{const}), then the set $\{U_0,U_1,V_0,V_1\}$, given
from the solution $\mu$ of the $\bar\partial$-problem in (\ref{ulax})
(\ref{vlax}), obeys the nonlinear evolution (\ref{main}).

The compatibility constraint (\ref{const}) will be seen to be essential.
At this stage it is important to note that the degree of complexity of this
constraint is usually greater than that of the nonlinear equation itself. Hence
the
main point to make here is that, in order to build from theorem 1 some {\em
practical} tools, we have to {\em make a decision}, for instance by choosing
an elementary solution of the constraint (\ref{const})

\vfill\eject

\setcounter{equation}{0}
\section{The nonlinear Schr\"odinger equation}

A natural choice for the arbitrary functions $\Lambda$ and $\Omega$,
for which the constraint (\ref{const}) is independent of the spectral transform
$R$,
is
\begin{equation}
\Lambda_t - \Omega_x + [\Omega, \Lambda]=0.\label{azer}\end{equation}

It is usefull to rewrite the theorem in this subcase:\\
{\bf Theorem 2}
\begin{center}
\fbox{\begin{minipage}{11.5cm}%
For
\begin{equation}
R_x = [R,\Lambda] + N , \hskip15pt R_t = [R, \Omega] + M,
\label{evolr}\end{equation}
\begin{equation}
\Lambda_t - \Omega_x + [\Omega, \Lambda]=0,\label{azero}\end{equation}
we have the nonlinear evolution equation
\begin{equation} U_{0,t}-V_{0,x}+[U_0,V_0]=
{\cal P}_k \{[V_0,U_1] - [U_0, V_1]\},\label{theo}\end{equation}
with
$$ U_0 = - {\cal P}_k\{\mu\Lambda\mu^{-1}\},\hskip10pt
U_1=\bar{\partial}^{-1}\{\mu (N-\bar{\partial}\Lambda)\mu^{-1}\},$$
$$ V_0=-{\cal P}_k\{\mu \Omega \mu^{-1}\},\hskip10pt
V_1=\bar{\partial}^{-1}\{\mu(M-\bar{\partial}\Omega)\mu^{-1}\},$$
if $M$ and $N$ satisfy the constraint
\begin{equation}N_t -[N,\Omega] - M_x + [M,\Lambda] = 0 .
\label{ntnx}\end{equation}
\end{minipage}}
\end{center}

The simplest non-trivial choice is realized by taking for $\Lambda$ and
$\Omega$ two {\em diagonal} matrices, {\em polynomial} in $k$ with {\em
constant coefficients}. In that case, the matrix $A$ given in
(\ref{adisp}) does vanish and a trivial solution to (\ref{const}) is
$M=N=0$.
Then we recover the usual hierarchies of integrable partial differential
equations. To set an example, let us chose
\begin{equation}
\Lambda=ik\sigma_3,\hskip10pt \Omega=ik^2\sigma_3,\hskip10pt M=N=0.
\label{dispnls}\end{equation}
Then from (\ref{ulax}) and (\ref{vlax})
\begin{equation}
U_1=0,\hskip10pt V_1=0 ,
\end{equation}
and from the Laurent series (\ref{laurent}) for $\mu(k)$
\begin{equation}
U_0=-ik\sigma_3+i[\sigma_3,\mu^{(1)}]\label{unls}\end{equation}
\begin{equation}
V_0=-i\sigma_3k^2+ik[\sigma_3,\mu^{(1)}]-i[\sigma_3,\mu^{(1)}]\mu^{(1)}
+i[\sigma_3,\mu^{(2)}]
\label{vnls}\end{equation}

The above two polynomials in $k$ are then inserted in the evolution equation
(\ref{theo}) and each coefficient of any power of $k$ is required to vanish.
Defining the anti-diagonal matrix $Q(x,t)$ by
\begin{equation} Q=i[\sigma_3,\mu^{(1)}],\label{qu}\end{equation}
we easily obtain: at order $k^2$ the equation (\ref{theo}) is automatically
verified, at order $k$ this equation gives
\begin{equation}i\left[\sigma_3,Q\mu^{(1)}-i[\sigma_3,\mu^{(2)}]\right]=
 Q_x.\label{mudeux}\end{equation}
The above equation can the be shown to be equivalent to the {\em diagonal} part
of the $k$-independent term in (\ref{theo}),
and it actually {\em defines} the quantity $i[\sigma_3,\mu^(2)]$ from $Q$.
Provided this information, the {\em anti-diagonal} part of (\ref{theo})
finally gives, again by using (\ref{mudeux}), the nonlinear Schr\"odinger
equation
\begin{equation}
i\sigma_3Q_t+{1\over2}Q_{xx}-Q^3=0.\label{matnls2}\end{equation}

At this point there is the important remark that $\Lambda$ and
$\Omega$ can be modified
in such a way as to keep the evolution equation unchanged (that is to
keep $U_0$ and $V_0$ unchanged) but to change the $(x,t)$-dependence of
$R(k,x,t)$. This will then generate, through the solution of (\ref{dbar}) and
of (\ref{qu}), a {\em larger class of solutions} of the NLS equation.
This can be done by choosing
\begin{equation}
{\cal P}_k\{\Lambda\}=ik\sigma_3,\hskip10pt {\cal P}_k\{\Omega\}
=ik^2\sigma_3,\label{spectralbound}
\end{equation}\begin{equation}
 N=\bar\partial\Lambda,\hskip10pt
 M=\bar\partial\Omega,
\label{nlsconst}\end{equation}
for which $U_0$ is still given by (\ref{unls}) and $V_0$ by (\ref{vnls}),
and for which $U_1=V_1=0$. Note that the constraint
(\ref{ntnx}) is automatically verified from (\ref{azero}) and (\ref{nlsconst}).

While $R(k,x,t)$ now obeys a much more general $(x,t)$-dependence,
the related nonlinear evolution is still the above NLS equation
(\ref{matnls2}),
and hence we have indeed here a method to generate a {\em larger class of
solutions}
to the NLS equation.
The method to perform this can be sketched as follows:\\
Step 1: get some explicit functions $\Lambda$ and $\Omega$ by means of
\begin{equation}
\left[\matrix{\Lambda_t-\Omega_x+[\Omega,\Lambda]=0\cr
{\cal P}_k\{\Lambda\}=ik\sigma_3,\hskip10pt {\cal P}_k\{\Omega\}
=ik^2\sigma_3}\right] \longrightarrow \{\Lambda(k,x,t),\Omega(k,x,t)\}.
\label {step1}\end{equation}
Step 2: solve the $(x,t)$-dependences for $R(k,x,t)$
\begin{equation}
\left[\matrix{R_x = [R,\Lambda] +\bar\partial\Lambda
 \cr R_t = [R, \Omega] + \bar\partial\Omega}\right]\longrightarrow
\{ R(k,x,t)\}.\label{step2}\end{equation}
Step 3: solve the linear Cauchy integral equation for $\mu(k,x,t)$
\begin{equation}
\left[\matrix{\mu(k) = {\bf 1} + {1\over 2i\pi} \int\!\!\!\int
{d\lambda\wedge d\bar{\lambda}\over \lambda - k} \mu (\lambda)
R(\lambda)}\right]\longrightarrow\{\mu(k,x,t)\}.\label{step3}\end{equation}
Step 4: compute the solution $Q(x,t)$ of the NLS equation by means of
$\mu(k,x,t)$
\begin{equation}
 Q(x,t)=i[\sigma_3,\mu^{(1)}(x,t)].\label{step4}\end{equation}

Steps 3 and 4 hereabove are the usual problems to be solved when a solution to
an
integrable nonlinear evolution is seeked and step 2 is a linear differential
equation which is solved through standard methods.
The novelty here lies in the step 1 which
solution actually reduces to solving the {\em nonlinear
integrable hierarchy} generated through the {\em potentials} $\Lambda$ and
$\Omega$
by the compatibility condition (\ref{azero}).
The problem is that of the stringent constraints
\begin{equation}\Lambda=ik\sigma_3+{\cal O}({1\over k}),\hskip15pt
\Omega=ik^2\sigma_3+{\cal O}({1\over k}).\label{stringent}\end{equation}
The technical complexity of the proposed method is such that we have not been
able to
find an explicit example. Note in particular that, when $\Lambda$ is not simply
$ik\sigma_3$, the scattering problem (the x-part of the Lax pair (\ref{lax}))
has to
be rebuilt completely. And this is a necessary step in order to find the {\em
constants of integration} for the two differential equations in (\ref{step2}).

In this context, the last point to check is the unicity of the solution of the
Cauchy
problem for the NLS eq. when $Q(x,0)$ is in the Schwartz space. This unicity
will hold
if, when analyzing $Q(x,0)$ through the Zakharov-Shabat spectral problem, in
other
words with
\begin{equation}\Lambda=ik\sigma_3,\end{equation}
we would obtain the {\em usual} time evolution of the spectral transform, in
other
words
\begin{equation}\Omega=ik^2\sigma_3.\end{equation}
This result is obtained here by first remarking that the compatibility equation
(\ref{azero}) in which $\Lambda=ik\sigma_3$ gives by induction (successive
powers of
$1/k$)
\begin{equation}\Omega^A=0.\label{omegaazero}\end{equation}
Second, we use the property given in the appendix that the spectral transform
$R$ is
an anti-diagonal matrix. Then the time evolution in (\ref{evolr}) can be
splitted in
its diagonal and anti-diagonal parts to give
\begin{equation}
\bar\partial\Omega^D=[\Omega^A,R],\hskip15pt
R_t=[R,\Omega^D]+\bar\partial\Omega^A.
\label{evolrda}\end{equation}
The first of the above equations can be explicitely integrated thanks to
(\ref{stringent}) and (\ref{omegaazero}) and we have
\begin{equation}\Omega^D=ik^2\sigma_3\end{equation}
which acheives the proof. Indeed, (\ref{evolrda}) now reads
\begin{equation}R_t=[R,ik^2\sigma_3].\end{equation}

\vfill\eject

\setcounter{equation}{0}
\section{Integrable asymptotic boundary value problems}

\subsection{Statement of the problem}

The general theorem 1 is shown here to serve as the basic tool for
demonstrating the following important result.

{\bf Theorem 3:}
\begin{center}
\fbox%
{\begin{minipage}{11.5cm}
The system of coupled equations for the three fields $q(x,t)$, $a_1(k,x,t)$ and
$a_2(k,x,t)$:
\begin{eqnarray}
q_t=  {\int_{-\infty}^{+\infty}} g dk\;a_1\bar a_2, \nonumber\\
a_{1,x}=qa_2,\label{siteq}\\
a_{2,x}-2ika_2=\sigma\bar q a_1\nonumber
\end{eqnarray}
(with $\sigma=\pm,\; x, k \in{\bf R},\; t>0$ and
$g=g(k)$ an arbitrary function in $L^2$ which may also depend on $t$),
is integrable for the {\em arbitrary asymptotic boundary values}
$a_j(k,\pm\infty,t)$.
\end{minipage}}\end{center}

More precisely we will give the evolution of the spectral transform in the
following
four cases:
\begin{equation}
a_1 {\displaystyle\mathop{\longrightarrow}_{x\rightarrow+\infty}}
I_1(k,t),\hskip10pt
a_2 {\displaystyle\mathop{\longrightarrow}_{x\rightarrow+\infty}}
I_2(k,t)e^{2ikx},
\label{casi}\end{equation}
\begin{equation}
a_1 {\displaystyle\mathop{\longrightarrow}_{x\rightarrow+\infty}}
J_1(k,t), \hskip10pt
a_2 {\displaystyle\mathop{\longrightarrow}_{x\rightarrow-\infty}}
J_2(k,t)e^{2ikx}
\label{casj}\end{equation}
\begin{equation}
a_1 {\displaystyle\mathop{\longrightarrow}_{x\rightarrow-\infty}}
K_1(k,t),\hskip10pt
a_2 {\displaystyle\mathop{\longrightarrow}_{x\rightarrow-\infty}}
K_2(k,t)e^{2ikx}
\label{cask}\end{equation}
\begin{equation}
a_1 {\displaystyle\mathop{\longrightarrow}_{x\rightarrow-\infty}}
L_1(k,t),\hskip10pt
a_2 {\displaystyle\mathop{\longrightarrow}_{x\rightarrow+\infty}}
L_2(k,t)e^{2ikx}.
\label{casl}\end{equation}
Here above the asymptotic boundary values ($I_1, I_2, J_1 ...$)
are arbitrary functions of the real variable $k$ in $L^2$
with an arbitrary dependence on the real parameter $t$.
Of course, due to the linearity of the equation for $a_j$ in (\ref{siteq}), any
of
the above behaviors can actually be obtained from a particular one. However we
will
see that it is quite usefull to have the four evolutions of the spectral
transform
separately as the link between these cases can well be very complicated. Note
also that to each of the above 4 cases there are two choices $\sigma=+$ and
$\sigma=-$ for which the corresponding evolutions of the spectral transform
will be
seen to be quite different.

The above system will be understood as the equation for the two envelopes $a_1$
and
$a_2$ of the two components of a high frequency field, interacting resonantly
with a
low frequency field of envelope $q$. The parameter $k$ actually measures a
frequency missmatch and $g(k)$ measures the relative intensity of the coupling
of
the waves for each value of $k$. Hence $g(k)$ has to be positive, symmetric
around
the proper resonance $k=0$ (even function of $k$) and of {\em finite support}.
Consequently the system (\ref{siteq}) is in three variables and is constituted
of a differential eq. in the variable $x$ running on the infinite line,
a differential eq. in the variable $t$ running on an arbitrary compact support
and
an integral eq. in the variable $k$ running on a given compact support (the
support of the measure $g(k)$). Given $g(k)$, $q(x,0)$ and the boundary values
for
the $a_j$'s, the system is closed and the problem of solving it is well posed.

The proof of the integrability of (\ref{siteq}) has been first given in
\cite{jld}\cite{jle} and
follows a large series of papers devoted to the {\em self-induced transparency}
(SIT)
equations of McCall and Hahn \cite{machan} which were given a Lax pair  and
soliton
solutions by Lamb \cite{lamb}. The sharp line case was then shown to be
completely
integrable in \cite{gibbon} where the N-soliton solution is displayed.
Then the inhomogeneously broadened case was also
shown to be completely integrable in \cite{abkane} where in particular
the evolution of the continuous part of the spectrum has been obtained, which
has
allowed the authors to demonstrate mathematically that this simple model does
contain
the property of transparency (the radiation was shown to vanish exponentially
as the
input laser pulse propagates in the medium).
Later, more general asymptotic
boundary conditions (a subcase of (\ref{casi}) for $|I_1|^2+|I_2|^2 =1$)
have been studied in \cite{gabi} and shown to lead to interesting physical
consequences. Simultaneously the scientific community became aware of the
importance
of this system as a fundamental model for resonant wave coupling processes in
the
{\em slowly varying amplitude approximation } (SVEA), role which is played by
the
nonlinear Schr\"odinger equation in the scalar case (no wave coupling). Indeed,
the system (\ref{siteq}) appears as a universal limit in many different
physical
systems, not only for the Maxwell-Bloch system in two-level one-dimensional
media (the
starting equations for SIT). It appears for instance in nonlinear optics to
describe
stimulated Raman scattering \cite{srs}, in diatomic chains of coupled harmonic
oscillators \cite{jldiat}, in plasma physics for stimulated Brillouin
scattering \cite{jlplas}... The point is that to each individual physical
situation,
there correspond different asymptotic boundary values \cite{jle}. Consequently,
the
integrability of (\ref{siteq}) has not a unique common implication on the
behaviour
of the solution and, as we shall see, the properties of the solution can well
be
drastically different from a problem to another.

The integrability of the system (\ref{siteq}) is proved in three steps. First
we
obtain a general integrable system of a structure similar to  (\ref{siteq})
for a particular choice of the entries $(N,\Lambda, M, \Omega)$ in theorem 1
(sec. 3). Second we consider some convenient reductions of the obtained
evolution equation. Third the values of the arbitrary functions occuring in the
chosen set $(N,\Lambda, M, \Omega)$ are determined in terms of the asymptotic
boundary values. The proof will actually only be sketched for the details can
be
found in \cite{jle}.

\subsection{ General integrable system}

Let us choose
\begin{equation}N=0,\hskip15pt \Lambda=ik\sigma_3,\label{zakspec}\end{equation}
\begin{equation}M=\left(\matrix{0&m^-(k,t)\cr m^+(k,t)&0}\right)
\exp[2ik\sigma_3x],
\label{Msing}\end{equation}
\begin{equation}
\bar\partial\Omega=p(k,t)\sigma_3,\hskip15pt
{\cal P}_k\{\Omega\}=0.\label{omegasing}\end{equation}
For that simple choice the compatibility constraint (\ref{const}) holds (note
that $A=0$) and the {\em potentials} $U$ and $V$ can be computed out of
(\ref{ulax})(\ref{vlax}) and read
\begin{equation}
U_0=-ik\sigma_3+i[\sigma_3,\mu^{(1)}],\hskip15pt U_1=0,
\label{zakprob}\end{equation}
\begin{equation}
V_0=0,\hskip15pt V_1=\bar\partial^{-1}\{\mu(M-\bar\partial\Omega)\mu^{-1}\}.
\label{auxprob}\end{equation}
The $2\times2$ matrix $\mu(k,x,t)$ obeys the differential equation (\ref{lax})
namely here:
\begin{equation}
\mu_x=-ik[\sigma_3,\mu]+\left(\matrix{0&q\cr r&0}\right)\mu,\label{zssp}
\end{equation}
where we have defined
\begin{equation}\left(\matrix{0&q\cr r&0}\right)=i[\sigma_3,\mu^{(1)}].
\label{qmu}\end{equation}

Finally the evolution equation (\ref{main}) of theorem 1 can be written
\begin{eqnarray}
q_t&=&-{1\over\pi}\int\!\!\!\int dk\wedge d\bar k
(2 \mu_{11}\mu_{12}\; p+
\mu_{11}^2 \;m^-\;e^{-2ikx}-\mu_{12}^2\;m^+\; e^{2ikx}),\nonumber\\
{ }\label{genersit}\\
r_t&=&-{1\over\pi}\int\!\!\!\int dk\wedge d\bar k
(2\mu_{21}\mu_{22}\; p +
\mu_{21}^2\;m^-\; e^{-2ikx}-\mu_{22}^2 \;m^+\;e^{2ikx}),
\nonumber\end{eqnarray}
where of course the $\mu_{ij}$'s are the matrix elements of $\mu(k,x,t)$.

The equations (\ref{zssp}) and (\ref{genersit}) constitute a system of 6
coupled
fields to which we have to associate the initial data ($q(x,0),\;\;r(x,0)$)
and some boundary values for the $\mu_{ij}$'s (for instance at $x=\infty$) for
all $t$.

\subsection{Constraints}

We shall work out three stages of simplifications of the system
(\ref{zssp}) (\ref{genersit}).
The  first one is a result of the constraint on the space of funtions to which
belongs the initial datum (we call it a {\em structural constraint}).
The second one is a constraint on the nature of the coupling term in the
evolution equation to be solved (we call it a {\em natural constraint}).
The last one is a {\em reduction} of the 6-field
system to a 3-field one by assuming a simple relation between $q$ and $r$
(we call it a {\em reduction constraint}).

{\bf i)} {\em\underline{ Structural constraints}}\\
To ensure consistency of the evolution (\ref{rt}) as an equation for {\em
distributions}, we must impose the constraints
which result from the support of the distribution $R(k,x,t)$ in the complex
$k$-plane.
This support depends on the space of functions to which belongs the initial
datum
$q(x,t)$, and, for $q$ in the Shwartz space, we recall in the appendix the main
results of the spectral theory for the Zakharov-Shabat spectral problem
(\ref{zssp}). There it is shown that the distribution $R$ has the following
structure
\begin{eqnarray}
R(k,x,t)=&{i\over2}
\left(\matrix{0&-\alpha^-(k,t)\delta^-(k_I)\cr
             \alpha^+(k,t)\delta^+(k_I)& 0} \right) e^{2ik\sigma_3
x}+\nonumber\\
{ }&+2\pi\sum_{n=1}^{N^\pm}
\left(\matrix{0&C_n^-(t) \delta(k-k_n^-)\cr
            C_n^+(t) \delta(k-k_n^+) & 0} \right) e^{2ik\sigma_3 x}
\label{structr}\end{eqnarray}
when we chose two basic solutions $\mu^\pm(k,x,t)$ of (\ref{zssp}) according to
the following behaviors for $k$ real
\begin{equation}
\left(\matrix{\beta^+&-e^{-2ikx}\alpha^-/\beta^-\cr
             0&{1/\beta^+}\cr}\right)
{\displaystyle\mathop{\longleftarrow}_{-\infty}}
\mu^+{\displaystyle\mathop{\longrightarrow}_{+\infty}}
\left(\matrix{1&0\cr e^{2ikx}\alpha^+&1\cr}\right)      \label{asympmuplus}
\end{equation}\begin{equation}
\left(\matrix{1/\beta^- & 0\cr
        -e^{2ikx}\alpha^+/\beta^+&\beta^-\cr}\right)
{\displaystyle\mathop{\longleftarrow}_{-\infty}}
\mu^-{\displaystyle\mathop{\longrightarrow}_{+\infty}}
\left(\matrix{1&e^{-2ikx}\alpha^-\cr
          0&1\cr}\right).
\label{asympmumoins}\end{equation}
It should be clear that the behaviors at $x=-\infty$ are actually
a consequence of the choice of the behaviors at $x=+\infty$.
Here above the new coefficients $\beta^\pm$ are explicitely defined from the
datum of $R$ in the appendix, we simply need here to remark that, from the
property (\ref{det}), we have the so called {\em unitarity relation}
\begin{equation}
\alpha^+\alpha^-+\beta^+\beta^-=1.
\label{unitarity}\end{equation}

The main constraint resulting from the above structure of the distribution $R$,
imposed on $M$ to ensure self consistency of the equation (\ref{rt}), read
\begin{equation}
m^\pm(k,t)=m_0^\pm(k,t)\delta^\pm(k_I)+
         \sum m_n^\pm(t)\delta(k-k_n^\pm).
\label{structm}\end{equation}
No such constraint applies on
$\Omega$ because it is a {\em fuction} multiplying $R$.

{\bf ii)} {\em\underline{ Natural constraints}}\\
Since we are interested on evolutions (\ref{siteq}) where the integral
on the r.h.s runs on the real axis only, it is readily seen on (\ref{genersit})
that it is necessary to set first
\begin{equation}
m_n^\pm(t)=0.\label{natredm}\end{equation}
Then it will be necessary that $p={\bar\partial}\omega$ be proportional to the
distributions
$\delta^\pm(k_I)$, hence
\begin{equation}
p(k,t)=p^+_0(k_R,t)\delta^+(k_I)+p^-_0(k_R,t)\delta^-(k_I).
\label{natredomeg}\end{equation}

Inserting now the set of constraints
(\ref{structm})(\ref{natredm})(\ref{natredomeg}) into
the evolution (\ref{genersit}), we obtain the following system
\begin{eqnarray}
q_t&=&{2i\over\pi}\int
2p^+_0\mu_{11}^+\mu_{12}^++2p^-_0\mu_{11}^-\mu_{12}^-+
m^-_0(\mu_{11}^-)^2 e^{-2ikx}-m_0^+(\mu_{12}^+)^2e^{2ikx},\nonumber\\
{ }\label{redgensit}\\
r_t&=&{2i\over\pi}\int
2p^+_0\mu_{21}^+\mu_{22}^++2p^-_0\mu_{21}^-\mu_{22}^-+
m^-_0(\mu_{21}^-)^2 e^{-2ikx}-m_0^+(\mu_{22}^+)^2e^{2ikx},
\nonumber\end{eqnarray}
where of course the matrix $\mu=\{\mu_{ij}\}$ solves (\ref{zssp}), and where
the
notation $\mu_{ij}^\pm$ is defined in (\ref{notepm}). It turns out (see
appendix)
that indeed the functions $\mu^\pm$ do obey the behaviors (\ref{asympmuplus})
(\ref{asympmumoins}).

{\bf iii)}  {\em \underline{Reduction constraints}}\\
It is convenient for physical applications to look for reduction of the above
6-field system to a simpler one (actually a 3-field system). This is done by
assuming the following {\em reduction}
\begin{equation}
r(x,t)=\sigma\bar q(x,t),\;\;\sigma=\pm\label{redq}\end{equation}
for which from (\ref{zssp})
\begin{equation}
\bar\mu_{11}^\pm=\mu_{22}^\mp,\hskip10pt \bar\mu_{12}^\pm=\sigma\mu_{21}^\mp,
\label{redmu}\end{equation}
and consequently using (\ref{structr}) in (\ref{start})
\begin{equation}
\alpha^+=\sigma\bar\alpha^-,\hskip10pt \beta^+=\bar\beta^-,\hskip10pt
k_n^+=\bar k_n^-,\hskip10pt C_n^+=\sigma\bar C_n^-.
 \label{redstructr}\end{equation}
The evolutions (\ref{rt}) together with (\ref{redgensit})
then imply  the following constraints
\begin{equation}
m^+_0(k,t)=\sigma\bar m^-_0(\bar k,t),\hskip10pt p^+_0(k,t)=-\bar p^-_0(\bar
k,t).
\label{redspec} \end{equation}
It is convenient to define lighter notations through
\begin{equation}
m_0= m_0^+,\hskip10pt p_0 =
p^+_0,\hskip10pt\alpha=\alpha^+,\hskip10pt\beta=\beta^+,
\hskip10pt C_n=C_n^+,\hskip10pt k_n=k_n^+.
\label{defpzero}\end{equation}

Finally, considering all the simplifications resulting from
the above three stages of constraints, the  system
(\ref{redgensit})(\ref{zssp}) reads
\begin{eqnarray}
q_t={2i\over\pi}{\int_{-\infty}^{+\infty}}
2p_0\mu_{11}^+\mu_{12}^+-2\bar p_0\mu_{11}^-\mu_{12}^-+
\sigma\bar m_0(\mu_{11}^-)^2 e^{-2i\lambda x}-
m_0(\mu_{12}^+)^2 e^{2i\lambda x}\nonumber\\
{  }\label{sitbasic}\\
\mu_{11,x}=q\mu_{21},\hskip10pt\mu_{21,x}-2ik\mu_{21}=\sigma\bar q
\mu_{11},\hskip15pt \mu_{12}^\pm=\sigma\bar\mu_{21}^\mp.
\nonumber\end{eqnarray}
Since the relevant datum for the evolution of $q(,t)$ is the function
$p_0(k,t)$, we need to compute $\omega(k,t)$ out of (\ref{natredomeg})
\begin{equation}
\omega(k,t)=-{1\over\pi}\int_{-\infty}^{+\infty}{d\lambda
\over\lambda-k+i0}\;p^+_0(\lambda,t)-{1\over\pi}\int_{-\infty}^{+\infty}
{d\lambda\over\lambda-k-i0}\; p^-_0(\lambda,t),
\label{omegapm}\end{equation}
that is, by using (\ref{redspec}), (\ref{defpzero}) and (\ref{ptem})
\begin{equation}
\omega(k,t)=-i(p_0+\bar p_0)-{1\over\pi}P\!\!\!\int{d\lambda
\over\lambda-k}(p_0-\bar p_0).\label{omegap}\end{equation}
It results that $\omega$ is purely imaginary which makes the constraints
(\ref{redstructr}) compatible with the evolution (\ref{rt}).

\subsection{ Evolution of the spectral transform}

To acheive the proof of integrability of the system (\ref{siteq}) with
arbitrary boundary values for the fields $(a_1, a_2)$, it is necessary to
find the transformation from (\ref{siteq}) to (\ref{sitbasic}) by
expanding the vector $(a_1, a_2)$ on a suitable basis constructed from the
vectors $(\mu_{11}^\pm, \mu_{21}^\pm)$. To do that it is enough to compare
their asymtotic boundary values (they both solve the same first order
differential linear system), and hence, from (\ref{asympmuplus})
(\ref{asympmumoins}) we can deduce for each of the cases (\ref{casi})
(\ref{casj}) (\ref{cask}) (\ref{casl}):
\begin{equation}
\left(\matrix{a_1\cr a_2}\right)=I_1 \mu_1^-
+I_2 \mu_2^+ e^{2ikx},
\label{relati}\end{equation}
\begin{equation}
\left(\matrix{a_1\cr a_2}\right)={K_1\over\beta^+}\mu_1^+
+{K_2\over\beta^-}\mu_2^- e^{2ikx},
\label{relatj}\end{equation}
\begin{equation}
\left(\matrix{a_1\cr a_2}\right)=J_1 \mu_1^+
+\beta^+J_2 \mu_2^+ e^{2ikx},
\label{relatk}\end{equation}
\begin{equation}
\left(\matrix{a_1\cr a_2}\right)={L_1\over\beta^+}\mu_1^+
+L_2 \mu_2^+ e^{2ikx}.
\label{relatl}\end{equation}

Next it is necessary to compute the product $a_1\bar a_2$ appearing in the
equation (\ref{siteq}) and to express it in terms of the only four products
appearing in (\ref{sitbasic}), namely
$\mu_{11}^+\mu_{12}^+$, $\mu_{11}^-\mu_{12}^-$,
$(\mu_{11}^-)^2 e^{-2ikx}$ and
$(\mu_{12}^+)^2 e^{2ikx}$. This is possible by repeatedly using the
Riemann-Hilbert problem for the function $\mu$, as shown in the appendix.
Actually we discover there that the real part of $p_0$ can be set to zero.
Once this is done, it is the sufficient to compare the equations (\ref{siteq})
and
(\ref{sitbasic}) and to equate the different coefficients of the above
mentionned terms. The result reads:

i) In the case (\ref{casi}) with (\ref{relati}):
\begin{eqnarray}
p_0&=&-i{\pi\over8}g
(\sigma|I_1|^2+|I_2|^2),\nonumber\\
m_0&=&{i\pi\over2}g[\sigma\bar I_1 I_2-{1\over2}\alpha
(\sigma|I_1|^2+|I_2|^2)]. \label{scatti}
\end{eqnarray}

ii) In the case (\ref{casj}) with (\ref{relatj}):
\begin{eqnarray}
p_0&=&-i{\pi\over8}g
[\sigma|J_1|^2(1+ \sigma|\alpha|^2 )+ |J_2|^2|\beta|^2+
J_1\bar J_2\alpha\bar\beta+\bar J_1 J_2\bar
\alpha\beta],\nonumber\\
{ }\label{scattj}\\
m_0&=&{i\pi\over2}g[\sigma\bar J_1 J_2\beta(1-{1\over2}
\sigma|\alpha|^2)- {1\over2}\bar\beta(\alpha)^2J_1\bar J_2+
{1\over2}\alpha|\beta|^2(\sigma|J_1|^2-
|J_2|^2)].\nonumber\end{eqnarray}

iii) In the case (\ref{cask}) with (\ref{relatk}):
\begin{eqnarray}
p_0&=&-i{\pi\over8}g
[(\sigma|K_1|^2+|K_2|^2) {1+\sigma|\alpha|^2\over1-\sigma|\alpha|^2}+
K_1\bar K_2{\alpha\over\beta^2}
     +\bar K_1 K_2{\bar\alpha\over\bar\beta^2}]\nonumber\\
{ }\label{scattk}\\
m_0&=&{i\pi\over2}g[\sigma\bar K_1 K_2\beta(\bar\beta)^{-1}
+{1\over2}\alpha
(\sigma|K_1|^2+|K_2|^2)].\nonumber\end{eqnarray}

iv) In the case (\ref{casl}) with (\ref{relatl}):
\begin{eqnarray}
p_0&=&-i{\pi\over8}g
[\sigma|L_1|^2({1+\sigma|\alpha|^2\over1-\sigma|\alpha|^2})+|L_2|^2+
L_1\bar L_2{\alpha\over\beta}
+\bar L_1 L_2{\bar\alpha\over\bar\beta}]\nonumber\\
{ }\label{scattl}\\
m_0&=&{i\pi\over2}g[{\sigma\over\bar\beta}(1-{1\over2}
\sigma|\alpha|^2)\bar L_1 L_2
- {1\over2}{\alpha^2\over\beta}L_1\bar L_2+
{1\over2}\alpha(\sigma|L_1|^2-|L_2|^2)].
\nonumber\end{eqnarray}

With these values in hand, the evolution of the spectral transform
defined in (\ref{structr}) is given by
\begin{eqnarray}
\alpha_t=2\omega\alpha-2im_0,\nonumber\\
k_{n,t}=0,\label{evolalpha}\\
C_{n,t}=2\omega(k_n)C_n,\nonumber\end{eqnarray}
where $\omega(k,t)$ is given from $p_0(k,t)$ in (\ref{omegap}).

It is convenient also, for future use, to obtain the evolutions of the
coefficient $\beta$ and of the quantity $|\alpha|^2$.
{}From the definition of $\beta^\pm$ given in the appendix, it is easy to
derive
\begin{equation}
\beta_t=-\beta{1\over2i\pi}\int{d\lambda\over\lambda-k-i0}\;\;
{(\sigma|\alpha|^2)_t\over1-\sigma|\alpha|^2}
\label{evolbeta}\end{equation}
and from the above evolution (\ref{evolalpha})
\begin{equation}
(|\alpha|^2)_t=2i(\bar m_0\alpha-m_0\bar\alpha).\label{evole}\end{equation}
Remember finally that, within the reduction (\ref{redq}), the unitarity
relation
(\ref{unitarity}) becomes
\begin{equation}
\sigma|\alpha|^2+|\beta|^2=1.\label{unitred}\end{equation}

\subsection{Comments}

We have shown here above that the time evolution (\ref{rt}) of the spectral
transform can be computed explicitely in terms of the asymptotic boundary
values
(\ref{casi})-(\ref{casl}) in the two cases $\sigma=\pm$, when $q(x,t)$ obeys
the
coupled system (\ref{siteq}). As can be seen in the preceding subsection, these
evolutions can be quite complicated and possibly {\em not explicitely
solvable}. In
such a case the property of {\em integrability} of the system  (\ref{siteq}) is
weakened by the constraints of the boundary values. Still the spectral
transform
remains a quite usefull tool as it maps the problem of the solution of
(\ref{siteq})
(3 variables with one on the infinite line) into the simpler problem of
solution of
(\ref{evolalpha}) (2 variables both on compact support).

We have displayed in \cite{jle} a few applications of the present results for
physical problems of resonant wave interaction. The important result to
emphasize
here is the strong dependance of the nature of the solution on the chosen
boundary
values. For instance, in the case of self-induced transparency, we obtain that
$|\alpha|^2\to0$ as $t\to\infty$ and this is precisely the {\em transparency
property}.
For stimulated Brillouin scattering in plasmas
we get $|\alpha|^2\to1$ as $t\to\infty$, and this is precisely the property of
{\em total reflexivity}.
We also give an example of special
boundary values in stimulated Raman scattering for which $|\alpha|^2\to\infty$
as $t\to t_s$ (finite), in that case the general solution blows up in finite
time.
There are other physical situations (under study now \cite{gino}) for which the
time evolutions (\ref{evolalpha}) (\ref{evolbeta}) and (\ref{evole}) are not
explicitely solvable. In that case, although the system possesses a Lax pair
and a
spectral transform, its integrability is partially destroyed by the constraints
due
to the boundary values. For instance, such a situation would be encountered
here if
considering the evolutions (\ref{evolalpha}) (\ref{evolbeta}) and (\ref{evole})
in the case (\ref{scattj}) with $J_1J_2\ne0$ and $J_i$ {\em external} (not
functionals of $\alpha,\beta...$).

As a last comment it is instructive to consider the case of self-induced
transparency which corresponds to the choice \cite{jle}
\begin{equation}
\sigma=-,\hskip10pt K_1=1,\hskip10pt K_2=0\label{choicesit}\end{equation}
in (\ref{scattk}). Hence we get
\begin{equation}
p_0=i{\pi\over8}g{1-|\alpha|^2\over1+|\alpha|^2},\hskip10pt
m_0=-i{\pi\over4}g\alpha,\label{paramsit}\end{equation}
and the evolution (\ref{evolalpha}) reads
\begin{equation}
\alpha_t=-{\pi\over2}g\alpha-{1\over2}\alpha P\!\!\!\int{d\lambda
g\over\lambda-k}
{1-|\alpha|^2\over1+|\alpha|^2}.\label{alphasit}\end{equation}
We have also from (\ref{evole})
\begin{equation}
(|\alpha|^2)_t=-\pi g|\alpha|^2,\label{evolesit}\end{equation}
which proves the {\em transparency}: $|\alpha|^2$ vanishes exponentially
as $t\to\infty$. The interesting (curious) fact is that, originally in
\cite{abkane}, the time evolution of the spectral transform has been written
for
the quantity
\begin{equation}\rho=\sigma\bar\alpha\beta(\bar\beta)^{-1},
\label{coeffleft}\end{equation}
which is, in the language of the scattering theory, the reflection coefficient
to
the left. It turns out that, by using successively (\ref{alphasit})
(\ref{evolesit}) and (\ref{evolbeta}), the evolution of $\rho(k,t)$
can be written with help of (\ref{ptem}):
\begin{equation}\rho_t=\rho{i\over2}\int{d\lambda\over\lambda-k+i0}\;\;g(\lambda).
\label{rhosit}\end{equation}
The point is that this evolution is {\em linear} while the equivalent one for
$\alpha$ (\ref{alphasit}) is not. This property lead the people thinking that
the
situation for coupled systems was much the same as that of scalar systems,
namely
that the difference concerns only the dispersion relation which from polynomial
becomes a singular function of $k$. This has produced a lot of works
\cite{lamb}\cite{gibbon}\cite{abkane}\cite{gabi}\cite{kaunew}\cite{melnik},
including those of the author \cite{jla}\cite{jlb}\cite{jlc}, where the
linearity
of the evolution of the spectral transform has been taken {\em a priori}, which
implicitely imposes constraints on the boundary values. We see here that the
linearity of the evolution can occur in particular situations, but is never a
general property. The first instance of a system of coupled waves corresponding
to a nonlinear evolution of the spectral transform has been given by KAUP in
\cite{caviton} in the case when the spectral problem is the Schr\"odinger
scattering problem. This problem has then been further studied in
\cite{kauplati}
and an extension to arbitrary boundary values is also feasible \cite{anouch}.
\vfill\eject

\setcounter{equation}{0}
\renewcommand{\theequation}{A.\arabic{equation}}
\section{Appendix: spectral and $\bar\partial$ problems}

\subsection{Spectral problem and basic solutions}

We briefly recall here the basic results on the
Zakharov-Shabat spectral problem which we write
for the $2\times2$ matrix $ \mu(k,x,t)$
\begin{equation}\mu_x+ik[\sigma_3,\mu]=Q\mu,\hskip10pt
Q=\left(\matrix{0&q(x,t)\cr
r(x,t)&0\cr}\right).
 \label{apzs}\end{equation}
Its two fundamental solutions, say $\mu^\pm$, are
determined by (a $t$-dependence is understood everywhere)
\begin{equation}\left(\matrix{\mu_{11}^+(k,x)\cr\mu_{21}^+(k,x)\cr}\right)=
\left(\matrix{1\cr0\cr}\right)+
\left(\matrix{-\int_{x}^{\infty}d\xi q(\xi)\mu_{21}^+(k,\xi)\cr
      \int_{-\infty}^{x}d\xi r(\xi)\mu_{11}^+(k,\xi)e^{2ik(x-\xi)}\cr}
       \right)                          \label{apint1}\end{equation}
\begin{equation}\left(\matrix{\mu_{12}^+(k,x)\cr\mu_{22}^+(k,x)\cr}\right)=
\left(\matrix{0\cr1\cr}\right)-
\left(\matrix{\int_{x}^{\infty}d\xi q(\xi)\mu_{22}^+(k,\xi)
                              e^{-2ik(x-\xi)}\cr
     \int_{x}^{\infty}d\xi r(\xi)\mu_{12}^+(k,\xi)\cr}
       \right)                        \label{apint2}\end{equation}

\begin{equation}\left(\matrix{\mu_{11}^-(k,x)\cr\mu_{21}^-(k,x)\cr}\right)=
\left(\matrix{1\cr0\cr}\right)-
\left(\matrix{\int_{x}^{\infty}d\xi q(\xi)\mu_{21}^-(k,\xi)\cr
     \int_{x}^{\infty}d\xi r(\xi)\mu_{11}^-(k,\xi)e^{2ik(x-\xi)}\cr}
       \right)                               \label{apint3}\end{equation}
\begin{equation}\left(\matrix{\mu_{12}^-(k,x)\cr\mu_{22}^-(k,x)\cr}\right)=
\left(\matrix{0\cr1\cr}\right)+
\left(\matrix{\int_{-\infty}^{x}d\xi q(\xi)\mu_{22}^-(k,\xi)
                              e^{-2ik(x-\xi)}\cr
     -\int_{x}^{\infty}d\xi r(\xi)\mu_{12}^-(k,\xi)\cr}
       \right)
\label{apint4}\end{equation}

The first column vector $\mu_1^+$ of the matrix $\mu^+$
is meromorphic in
${\hbox{Im }}(k)>0$ where it has a finite number $N^+$ of poles $k_n^+$
(assumed
to be simple). The second vector $\mu_2^+$ is holomorphic in ${\hbox{Im
}}(k)>0$.
The vector $\mu_1^-$ is holomorphic in ${\hbox{Im }}(k)<0$ while the second one
$\mu_2^-$ is meromorphic in ${\hbox{Im }}(k)<0$ where it has a finite number
$N^-$ of pole $k_n^-$ (simple).
One can check directly on the above integral equations that we have the
relations
\begin{equation}\displaystyle\mathop{\rm Res}
_{k=k_n^+}\mu_1^+(k)=iC_n^+\mu_2^+(k_n^+),
                                           \label{apres1}\end{equation}
\begin{equation}\displaystyle\mathop{\rm Res}_{k=k_n^-}
\mu_2^-(k)=-iC_n^-\mu_1^-(k_n^-),
                                              \label{apres2}\end{equation}
which define the {\em normalization coefficients} $C_n^\pm$.

The function $\mu(k)$ defined as $\mu^+$
in the upper half plane and $\mu^-$ in the lower is then discontinuous
on the real $k$-axis. Its discontinuity can be expressed simply in terms
of $\mu$ itself as
\begin{equation}\mu_1^+ - \mu_1^- = e^{2ikx}\alpha^+\mu_2^+,
\label{aprh1}\end{equation}
\begin{equation}\mu_2^+ - \mu_2^- =  -e^{-2ikx}\alpha^-\mu_1^-,
\label{aprh2}\end{equation}
which define the
{\em reflection coefficients} $\alpha^\pm(k)$:
\begin{equation}\alpha^+(k)={\int_{-\infty}^{+\infty}} d\xi\;r(\xi)\;\mu_{11}^+
(k,\xi)e^{-2ik\xi},\label{aprefl1}\end{equation}
\begin{equation}\alpha^-(k)={\int_{-\infty}^{+\infty}} d\xi\;q(\xi)\;\mu_{22}^-
(k,\xi)e^{2ik\xi}.\label{aprefl2}\end{equation}
We can also define the transmission coefficients $\beta^\pm$
\begin{equation}\beta^+(k)=1-{\int_{-\infty}^{+\infty}}
d\xi\;q(\xi)\;\mu_{21}^+
(k,\xi),\label{aptrans1}\end{equation}
\begin{equation}\beta^-(k)=1-{\int_{-\infty}^{+\infty}} d\xi\;r(\xi)\;
\mu_{12}^-(k,\xi).\label{aptrans2}\end{equation}
The integral equations (\ref{apint1}) to (\ref{apint4}) give
the following behaviors at large $x$:
\begin{equation}\left(\matrix{\beta^+&-e^{-2ikx}\alpha^-/\beta^-\cr
             0&{1/\beta^+}\cr}\right)
{\displaystyle\mathop{\longleftarrow}_{-\infty}}
\mu^+{\displaystyle\mathop{\longrightarrow}_{+\infty}}
\left(\matrix{1&0\cr e^{2ikx}\alpha^+&1\cr}\right)
\label{apbehav1}\end{equation}
\begin{equation}\left(\matrix{1/\beta^- & 0\cr
        -e^{2ikx}\alpha^+/\beta^+&\beta^-\cr}\right)
{\displaystyle\mathop{\longleftarrow}_{-\infty}}
\mu^-{\displaystyle\mathop{\longrightarrow}_{+\infty}}
\left(\matrix{1&e^{-2ikx}\alpha^-\cr
          0&1\cr}\right)                         \label{apbehav2}\end{equation}
where we have used
that from (\ref{apzs}), det$(\mu)$ is $x$-independent. Consequently
we have also the unitarity relation
\begin{equation}\alpha^+\alpha^-+\beta^+\beta^-=1.
\label{apunit}\end{equation}
To obtain the asymptotic behaviors  $-e^{-2ikx}\alpha^-/\beta^-$ of
$\mu_{12}^+$ and $e^{-2ikx}\alpha^+/\beta^+$ of $\mu_{21}^-$ as $x$
goes to
$-\infty$, one must use the relations (\ref{aprh1}) and (\ref{aprh2})
repeatedly in
(\ref{apint1}) and (\ref{apint3})  respectively.

Solving the direct scattering problem consists in solving for given
$Q(x)$ the integral equations (\ref{apint1}) to (\ref{apint4})
and then calculating the {\em spectral data} $\cal S$:
\begin{equation}{\cal S}=\{\alpha^\pm(k), k\in{\bf R}; k_n^\pm, C_n^\pm,
n=1,..,N^
\pm\}.\label{apsd}\end{equation}

\subsection{$\bar\partial$-problem and spectral transform}

Since the work  \cite{beals}, we know that the
solution of the inverse problem, i.e. the reconstruction of $Q$ from
$\cal S$, is given by a Cauchy-Green integral equation which solves a
${\bar\partial}$-problem for the function $\mu(k)$ previously defined.
Actually the
${\bar\partial}$-problem is simply the formula which summarizes  the
analytical properties of $\mu(k)$, it reads
\begin{equation}{\frac{\partial}{\partial\bar k}}\mu(k)=\mu(k)R(k)
\label{apdbar}\end{equation}
where the distribution $R(k)$ is the {\em spectral transform} and is
given from the spectral data $\cal S$ by
$$R(k)={i\over2}
\left(\matrix{0&-\alpha^-(k)\delta^-(k_I)\cr
             \alpha^+(k)\delta^+(k_I)& 0} \right) e^{2ik\sigma_3 x}+$$
\begin{equation}+2\pi\sum_{n=1}^{N^\pm}
\left(\matrix{0&C_n^- \delta(k-k_n^-)\cr
            C_n^+ \delta(k-k_n^+) & 0} \right) e^{2ik\sigma_3 x}.
 \label{apstructr}\end{equation}
This formula can be demonstrated simply by noting that
\begin{equation}\begin{array}{lcl}
k\in{\bf R}:&{\bar\partial}\mu(k)=&{i\over2}\mu(k)[\delta^+(k_I)-
\delta^-(k_I)]\\ {  }&{  }&{  }\\
{\hbox{Im }}(k)>0:&{\bar\partial}\mu(k)=&\left(-2i\pi\sum\delta(k-k_n^+)
\displaystyle\mathop{\rm Res}_{k_n^+}\mu_1^+(k)\;,\;0\right)\\
{  }&{  }&{  }\\
{\hbox{Im }}(k)<0:&{\bar\partial}\mu(k)=&\left(0\;,\;2i\pi\sum
\delta(k-k_n^-)
\displaystyle\mathop{\rm Res}_{k_n^-}\mu_2^-(k)  \right).
\end{array}                                     \end{equation}
The above ${\bar\partial}$-problem is completed by the asymptotic
behavior of $\mu$
at large $k$, which is obtained from (\ref{apint1}) to (\ref{apint4}) by
integration
by parts and reads
\begin{equation}k\rightarrow\infty\;\Rightarrow\;\mu(k)={\bf 1}+{\cal O}\left(
\frac{1}{k}\right).\end{equation}
Finally the solution of the equation (\ref{apdbar}) obeying the above behavior
is obtained by solving the following Cauchy-Green integral equation:
\begin{equation}\mu(k)={\bf 1}+{1\over2i\pi}\int\!\!\!\int
\frac{d\lambda\wedge d\bar\lambda}{\lambda-k}
\;\;\mu(\lambda)R(\lambda).\label{apcauchy}\end{equation}

By comparaison of the different powers of $1/k$ in the equation (\ref{apzs}),
we readily obtain
\begin{equation}Q=i[\sigma_3,\mu^{(1)}],      \end{equation}
where we have defined $\mu^{(1)}$ through
\begin{equation}\mu(k)={\bf 1}+\sum_{1}^{\infty}k^{-n}\mu^{(n)}. \end{equation}
Therefore the inverse problem is solved by the
integral equation (\ref{apcauchy}).
\vfill\eject

\end{document}